Short Paper*

# Narrowband-IoT (NB-IoT) and IoT Use Cases in Universities, Campuses, and Educational Institutions: A Research Analysis

Lyberius Ennio F. Taruc
Graduate School, Polytechnic University of the Philippines, Philippines
Computer Engineering Department, FEU Institute of Technology, Philippines
lyberiusennioftaruc@iskolarngbayan.pup.edu.ph
lftaruc@feutech.edu.ph
ORCID: 0009-0008-3566-5666
(corresponding author)

Arvin R. De La Cruz
Graduate School, Polytechnic University of the Philippines, Philippines
ORCID: 0000-0001-7325-5301
docardelacruz@gmail.com

## Abstract

*Purpose* – The main objective of this research paper is to analyze the available use cases of Narrowband-IoT and IoT in universities, campuses, and educational institutions.

*Method of Research* – A literature review was conducted using multiple databases such as IEEE Xplore, ACM Digital Library, and Scopus.

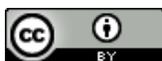

*Results* – The study explores the benefits of IoT adoption in higher education. Various use cases of NB-IoT in educational institutions were analyzed, including smart campus management, asset tracking, monitoring, and safety and security systems. Of the six use cases assessed, three focused on the deployment of IoT Things, while three focused on NB-IoT Connectivity.

*Conclusion* – The research paper concludes that NB-IoT technology has significant potential to enhance various aspects of educational institutions, from smart campus management to improving safety and security systems.

*Recommendations* – The study recommends further exploration and implementation of NB-IoT technology in educational settings to improve efficiency, security, and overall campus management.

*Research Implications* – The research highlights the potential applications of NB-IoT in universities and educational institutions, paving the way for future studies in this area.

*Social Implications* – The social implications of this research could involve enhancing the overall learning experience for students, improving campus safety, and promoting technological advancements in educational settings.

## INTRODUCTION

Educational institutions--universities, colleges, and campuses--have been increasingly embracing digitalization and integrating technology into their day-to-day operations. (Yasmin et al., 2020) This shift towards digitalization has been driven by various factors, including the need for improved efficiency, enhanced learning experiences, and the desire to keep up with the ever-evolving technological landscape. (Li et al., 2018)

In recent years, there has been a growing interest in applying Internet-of-Things (IoT) to existing processes and methods in academia. Researchers and academics have recognized the potential benefits that IoT can bring, more specifically, the deployment and use of Narrowband IoT (Li et al., 2018).

NB-IoT promises to bring the following benefits: enhanced connectivity, improved energy efficiency, increased device scalability, and extended range of coverage. Though generic, these benefits can also be extended toward an academic setting. According to research, NB-IoT offers several advantages for the development and implementation of IoT in universities, campuses, and educational institutions (Gbadamosi et al., 2020), and these can be reaped based on an institution's specific requirements.

## OBJECTIVE

The main objective of this research paper is to analyze the available use cases of Narrowband-IoT and IoT in universities, campuses, and educational institutions.

## RESEARCH SCOPE

This research focused specifically on the applications and potential use cases of NB-IoT in universities, campuses, and educational institutions. Several applications of NB-IoT within the context of academia may include smart campus management, asset tracking, environmental monitoring, and safety and security systems. (Guo & Nazir, 2021)

Analysis was done based on an institution's available use case scenario, at least one per institution, but may have more depending on availability. Any personally identifiable information that is irrelevant to the study was excluded according to relevance and data privacy.

## METHODOLOGY

A systematic literature review was conducted to analyze the use cases of Narrowband Internet of Things (NB-IoT) technology within universities, campuses, and educational institutions. The review covered publications in IEEE Xplore, ACM Digital Library, and Scopus from 2013 to 2023. Search terms included "Narrowband IoT", "Internet-of-Things", and "NB-IoT" combined with terms related to educational settings and academia, such as "Education", "Teaching", "Smart Campus" and "Smart University". Inclusion criteria focused on peer-reviewed English publications detailing NB-IoT implementations or potential use cases within the educational context. Exclusion criteria removed publications lacking relevance to NB-IoT or the educational sector, those without sufficient detail, and non-English or inaccessible publications.

Following title/abstract and full-text screening, seven publications were included in the final analysis. Data extraction involved documenting the study context, NB-IoT applications, technologies, outcomes, country of origin, and the publication year. A simple criterion for comparing the seven use cases was developed, according to the architectural elements found in a telecommunications network, of which IoT can be associated with: 1. Terminal Point or User Equipment (UE), the end device closest to the user or subscriber; 2. Connectivity, or the technology used for network connection; 3. Application Host, or the Service being availed by the by the UE; and 4. Focus of Study, or the research paper's focus on conducting the study.

The findings highlight use cases and identify areas for future research, including scalability, cost-effectiveness, and integration with other emerging technologies. While the review aimed for comprehensiveness, it acknowledges potential limitations in database and search term selection.

# NB-IoT: THE CURRENT STATE OF TECHNOLOGY

## *Definition of Narrowband IoT*

Narrowband IoT (NB-IoT) is a cellular technology that is designed to connect low-power, battery-operated devices to the internet. It is "a wireless telecommunications technology standard developed by 3GPP... [which] uses the same sub-6 GHz wireless spectrum as the 4G LTE technology, but unlike 4G LTE and other previous wireless telecommunications standards, NB-IoT (along with LTE-M) was developed with the IoT in mind." (Vos, n.d.)

Essentially, the goal of NB-IoT is to bring connectivity to Things using the cellular network, specifically mobile wireless broadband, as a primary medium.

It shares similarities to Long Term Evolution Machine Type Communication, or LTE-M / LTE-MTC, with the latter utilizing the 4G LTE network, hence producing average data rates of up to 1Mbps, unlike in NB-IoT which can only produce up to 200kbps. (Labs, n.d.)

Both NB-IoT and LTE-M are tailored to cater to IoT applications that do not necessitate high data speeds but do demand devices that meet specific criteria: affordable; capable of operating for over ten years solely on battery power; able to connect to cellular networks in crowded environments; and accessible in isolated rural areas, indoors, or even underground.

This study primarily focused on NB-IoT and its relevant use cases in IoT, although there are various bases for comparing different technologies, within the context of the Education sector and Academia.

## *Architecture*

The NB-IoT Architecture contains the following components: the NB-IoT device, the end device that connects to the NB-IoT network (Figure 1). It can be a sensor, actuator, or other IoT device; the NB-IoT base station (eNB), the radio access network (RAN) node that connects the NB-IoT devices to the core network; the Core network, which consists of the Evolved Packet Core (EPC) and the Serving Gateway (SGW). The EPC provides the core network functions, such as routing, security, and mobility management. The SGW is responsible for tunneling data between the RAN and the EPC; the Packet Data Gateway (PGW), or the gateway between the core network and the Internet. It is also responsible for routing data between the IoT devices and the applications that they are connected to; and the Application server which hosts the applications that the IoT devices are connected to.

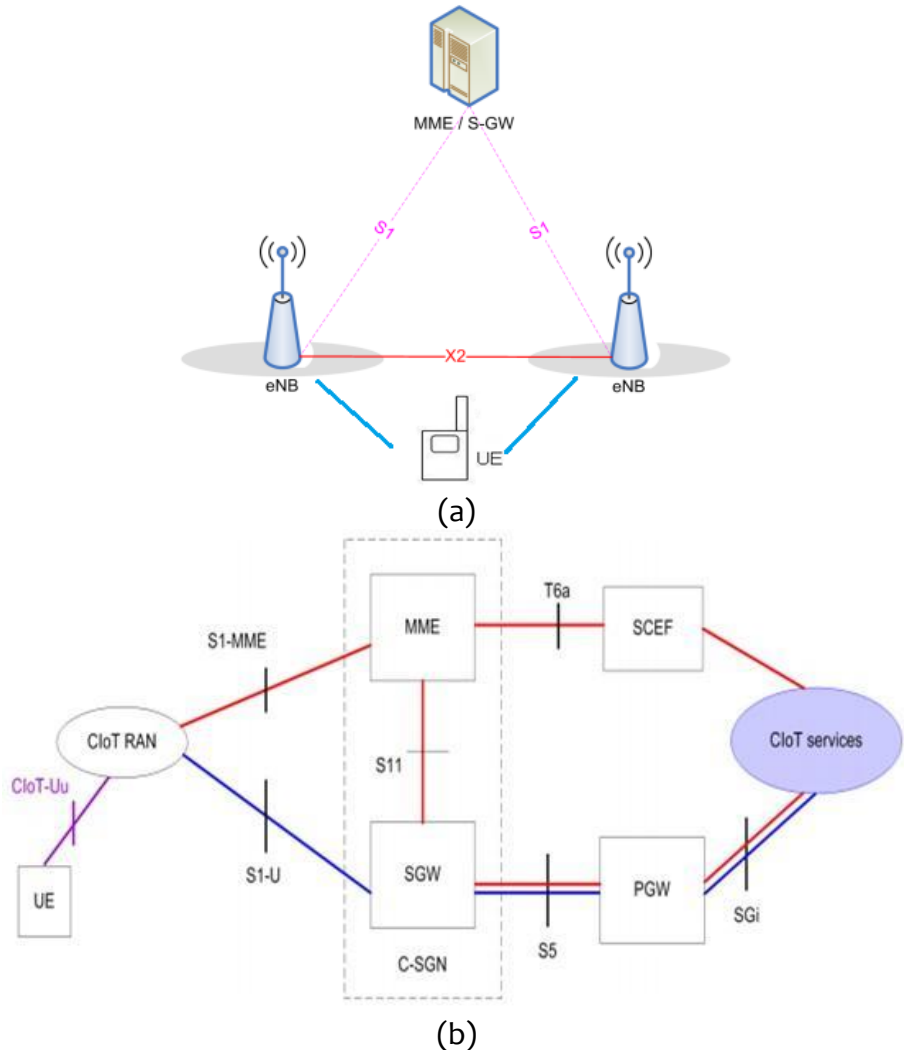

Figure 1. The general architecture of the NB-IoT Access (b) and Core (b) Networks.

This architecture is based on the LTE (Long Term Evolution) standard but has been optimized for low power and wide area coverage (Popli et al., 2019). The devices, or Things, that connect to the network can operate in deep sleep mode for most of the time and only wake up to transmit or receive data. The architecture is also designed to provide wide area coverage, as devices can connect to the network even in areas with poor cellular coverage. This makes NB-IoT ideal for applications such as smart metering, asset tracking, and environmental monitoring (RF & Wireless Vendors and Resources, n.d.). Regardless of which use case it applies, NB-IoT shares the common goal of providing long-range connectivity with low energy consumption via a network that can support many low-throughput devices.

# LITERATURE AND USE CASE REVIEW

## *Role and Application of NB-IoT in the Education Sector*

In a smart campus, NB-IoT can be used for a variety of applications: Asset tracking or the tracking of the location of assets, such as equipment, vehicles, and personnel; Security and safety for the improvement of security and safety on campus; Education, Teaching, and Learning Management to improve education or as a means of teaching; and, Campus Navigation System or the development of a centralized campus navigation system for aiding faculty, staff, students, and visitors in reaching their destination.

By enabling a wide range of applications that improve efficiency, sustainability, security, and education, NB-IoT can help to create a more connected and intelligent campus environment (Lin et al., 2018). The following use cases are then detailed.

## *Asset Tracking: A Low-Power Asset Tracking System Using Narrowband IoT— Telkom University, Indonesia*

The study analyzed the performance of a telemetry system using HTTP and a NB-IoT modem and SIM card. The data was transferred via HTTP, with the body request containing modem current usage, geolocation data, connectivity status, and miscellaneous data. The total request length ranged from 957B to 970B. The HTTP throughput was measured based on HTTP payload size and the time it took to complete the request. The NB-IoT modem and SIM card were connected via UART, supporting GPRS, NB-IoT, and LTE-CATM. The system also had a power saving mode (PSM) and an EDRX cycle of 122.8 seconds (Figure 2).

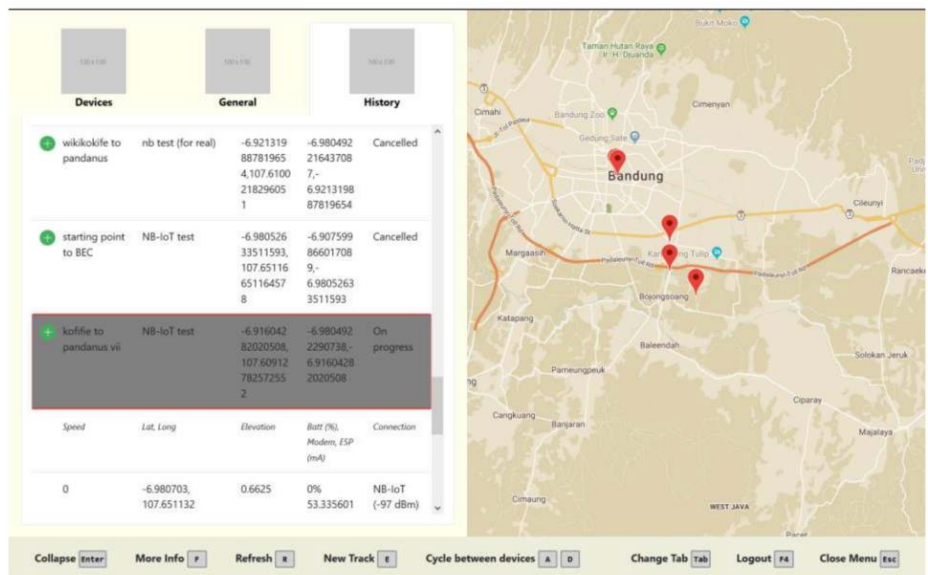

*Figure 2.* Asset Tracking Web UI of the NB-IoT system, showing the graphical locations of the Things around Bandung.

The ESP32 microcontroller was used to generate HTTP requests and record power consumption data. A Windows-based tablet was connected to the microcontroller to gather current data and provide a debugging interface, and power for the modem with other onboard peripherals. The results showed that NB-IoT is not suitable for HTTP applications or TCP with a large payload. Alternative transport methods, such as UDP, Protobuf, and JWT authentication, could improve connectivity throughput and reliability. The NB-IoT system underperformed compared to GPRS due to poor signal quality, payload, and transport methods (Bima et al., 2020).

### *Education, Teaching, and Learning Management: Impact of Smart Classrooms in Educational Institutions using IoT—SEGi University, Malaysia*

The study explored the benefits of Internet of Things (IoT) adoption in higher education and its implications. It involved 52 students from India and Malaysia, focusing on smart surroundings, smart learning methodologies, and smart participants. The research hypotheses include students' understanding of IoT, its adoption in higher education programs, its impact on intranet and extranet connectivity, enhanced access to learning resources, and improved student performance. The study concluded that higher education institutions should enhance their IoT infrastructure to meet global requirements. IoT-based smart classrooms are needed to develop efficient and knowledgeable learners without losing precious time to cope with growing digital technology. (Preethy & Parthasarathy, 2021)

### *Education, Teaching, and Learning Management: NB-IoT Micro-Operator for Smart Campus: Performance and Lessons Learned in 5GTN—University of Oulu, Finland*

The paper explored the deployment and performance evaluation of an NB-IoT micro-operator within the 5G Test Network (5GTN) for a smart campus at the University of Oulu. The study investigated the connectivity infrastructure provided by the micro-operator, focusing on NB-IoT technology's suitability for diverse machine-based applications indoors. Through extensive testing, the research demonstrates good indoor coverage, application-layer latency in the order of seconds, varying uplink and downlink throughputs, and the dynamics of transmit power control affecting energy consumption. The findings highlighted the viability of NB-IoT for non-critical machine applications indoors and emphasized the importance of micro-operators in enhancing connectivity performance. Recommendations included further research on network parameters' impact on performance and addressing energy efficiency and scalability for future deployments. (Yasmin et al., 2020)

*Security and Safety: Developing and Validation of a Smart Campus Safety and Security Framework at NICTM Using Internet of Things (IoT)— National Institute of Construction Technology and Management (NICTM), Nigeria*

The National Institute of Construction Technology and Management (NICTM) in Nigeria developed a smart campus safety and security framework using the Internet of Things (IoT). The framework used embedded system cameras to detect intruders using passive infrared (PIR) motion sensors. The information is sent to the premise owner via a telegram bot, allowing them to request general conditions from a remote location. The technology was validated in real-world scenarios and the feedback data was analyzed for potential intruders.

This study proposed an IoT-based safety and security architecture for the NICTM campus in Uromi, using passive infrared motion sensors, Figure 3. The system detects unauthorized intruders, uses proximity sensors, and sends photos to the premise owner via a Telegram bot. The system is based on bi-directional communication and one-way communication. C++ programming language was used for the design and validation.

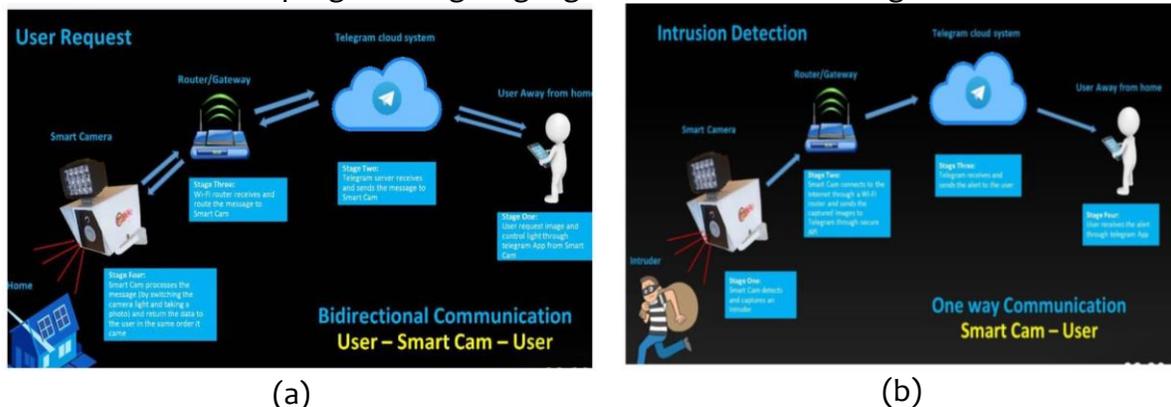

(a)                 (b)

*Figure 3.* Architecture (a) and Intrusion Detection process flow (b) of the Security System used in the NICTM campus in Uromi, Nigeria.

The system showed a good-quality display of photos taken in real-time, but poor-quality feedback was observed when unauthorized entry was initiated at night (Figure 4). The study recommended future work to design the camera in a way that is not visible to the would-be offender or intruder, and to expand the project by providing a big database for video recording and uploading information for crowd directing (Akpojedje & Efe, 2023).

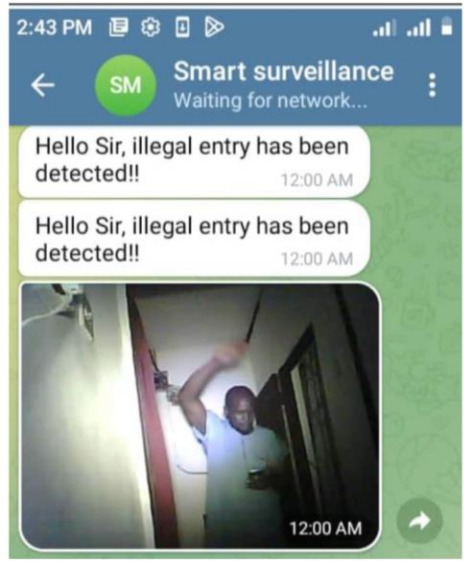

*Figure 4.* Sample photo snapshot and messaging system log sent via Telegram.

## *Campus Navigation: Smart Campus Navigation System— Shree L. R. Tiwari College of Engineering, India*

The project is an implementation of an Internet of Things (IoT) smart campus model that includes application services tailored for campus navigation use. Outdoor locations can be monitored by map apps like Google Maps, GPS navigation, and offline GPS maps.

The campus navigation system being suggested utilized RFID cards, RFID reader data, and servo motors to monitor location in outdoor and indoor settings. The door lock system is created with an Arduino UNO, servo motor, breadboard, jumper wires, and RF ID sensor. The RFID sensor processed user input and authenticated it with the Arduino Uno. When the input is deemed genuine, the microcontroller transmits the output to the servo motor, causing it to rotate 180 degrees. Figure 5 depicts the theoretical framework of the system.

This approach offered an accurate route through a campus or institute, assisting individuals who are not familiar with the university grounds in navigating without difficulty. University campuses with extensive grounds and vast facilities require it. The system was designed and implemented using NodeMcu, cloud technology, and the Blynk app. (Amisha et al., 2023)

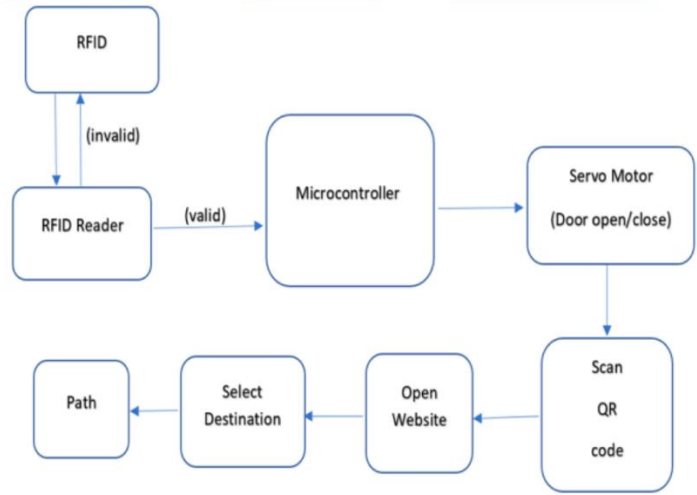

*Figure* 5. System's Theoretical Framework.

## Connectivity: Coverage Analysis of NB-IoT and Sigfox: Two Estonian University Campuses as a Case Study—University of Tartu and Tallinn University of Technology, Estonia

The paper investigated the coverage analysis of NB-IoT and Sigfox in Estonian university campuses, specifically in Tartu and Tallinn. By deploying nodes in various locations within the campuses, the study evaluates the performance of these LPWAN technologies in outdoor, indoor, and deep-indoor scenarios. Parameters such as RSSI, RSRP, and RSRQ are considered to assess network connectivity and reliability. Test results showed that both NB-IoT and Sigfox exhibit good coverage in outdoor environments with minimal packet losses. However, in indoor settings, NB-IoT demonstrated higher reliability compared to Sigfox, possibly due to re-transmissions. In deep-indoor or underground scenarios, coverage outages were observed for NB-IoT in Tartu, indicating weaker performance in that city. Figure 6 shows examples of the NB-IoT and Sigfox nodes (a) and deployment scenarios (b).

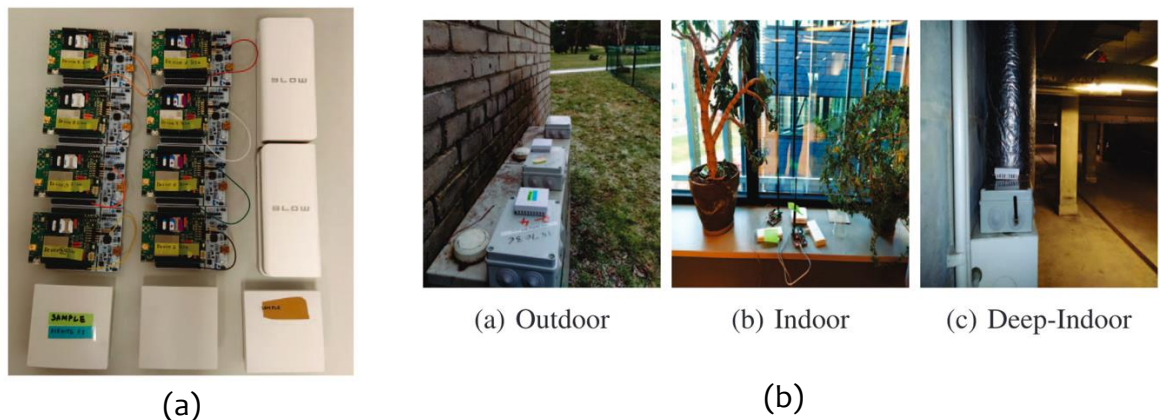

*Figure 6.* NB-IoT and Sigfox nodes (a) and deployment scenarios (b) used in the Estonian University Use Case.

The study contributed to the understanding of LPWAN technologies by providing real-time coverage analysis in different propagation conditions. The research methodology included deploying nodes programmed to transmit data to base stations, which are then forwarded to a backend system for analysis. The paper also acknowledged the support received from LPWAN service providers and funding agencies for the study. The findings suggest that NB-IoT and Sigfox offer promising performance in outdoor and indoor scenarios, with NB-IoT showing higher reliability in indoor environments. The results served as a reference for operators and researchers in evaluating the coverage and signal quality of LPWAN technologies in similar deployment settings. (Poddar et al., 2020)

## Campus Logistics: IoT-based Automatic Attendance Management System— Siddaganga Institute of Technology, India

This paper proposed an intelligent method for managing attendance using RFID and Internet of Things (IoT) technology to automate attendance management in educational institutions (Figure 7). The study focused on utilizing Raspberry Pi 3 and RFID sensors to develop an efficient and effective system for recording and monitoring student attendance. Aside from the physical device, an accompanying Android app with XAMPP was developed to serve as the system's backend (Sri Madhu et al., 2017). While RFID was used by the system as its frontend communication, a local network on-premises might have been used with the backend.

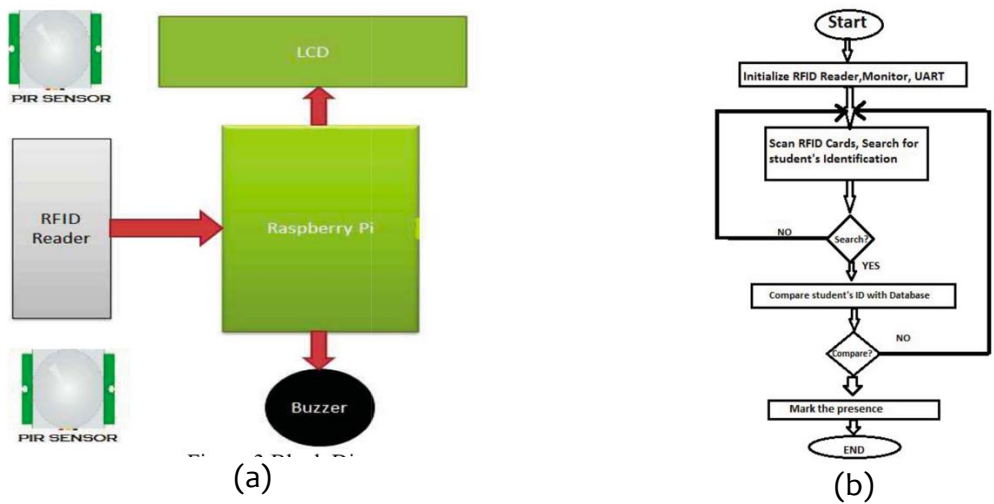

(a)          (b)

*Figure 7.* Attendance Monitoring System Theoretical Framework (a) and System Flow Chart (b).

## FINDINGS AND USE CASE SYNTHESIS

Assessing each use case's architecture, they share the common elements of what an IoT-based deployment should contain: (1) an IoT-enabled device that serves as the end-user's terminal point, or UE; (2) a means of connectivity, on most of the cases the utilization of the mobile network or broadband network unless otherwise disclosed; and (3) the application host or server which either provides a service or collects data from the UE.

Of the seven use cases of IoT, only six can be assessed in terms of provided network architecture, as the study made by SEGi University used a survey as its research instrument and no network architecture was designed. Regardless, it was still included due to its relevance to IoT within the academe.

Only three of the six use cases included NB-IoT as a key technology. The other three mentioned IoT but focused on the development of the end-device or IoT Thing. Though unspecified, the use cases from NICTM, SLRTCE, and SIT have internet as their minimum requirement, as IoT requires internet connectivity for data-on-demand transmissions (Gubbi et al., 2013).

Table 1. Summary of Network Architecture Elements per Use Case

| System Name | Terminal Point (UE) | Connectivity | Application Host (Service) | Focus of Study |
|---|---|---|---|---|
| **Telkom University** | ESP32 MCU | NB-IoT | Modem | Connectivity |
| **Oulu** | kDC-5737 NB-IoT Knowyou Nodes (Proprietary) | NB-IoT | EPC | Connectivity |
| **NICTM** | Webcam + In-house PCB | Unspecified | Telegram | IoT Thing |
| **SLRTCE** | Arduino Uno | Unspecified | School System | IoT Thing |
| **UT+TallTech** | NB-IoT DORM Nodes + Sigfox Airwits Nodes (Proprietary) | NB-IoT | End-to-End | Connectivity |
| **SIT** | RaspberryPi 3 | Unspecified | School System | IoT Thing |

A point to consider with the three NB-IoT use cases is the presence and general availability of a dedicated NB-IoT network within their locale: Telkom University and the UT+TallTech collaboration each used their local operators' infrastructure, while Oulu developed their infrastructure in-house. This aligns with the architectural requirements of NB-IoT which is derived from Legacy LTE (Rastogi et al., 2020).

# CONCLUSION

The implementation of a Smart Campus using NB-IoT is a method that is both dependable and efficient, and it may assist educational institutions in enhancing their levels of efficiency, sustainability, and safety. Despite its promises and numerous implementations, some of which have been stated in this paper, additional development is still needed to optimize its practicality and scope of availability.

In addition to the benefits of NB-IoT technology in managing various aspects of a smart campus, other technologies can also be integrated and included that can contribute to its overall success as a use case in academia. For instance, there is ongoing research on using augmented reality applications to provide contextual information and guide users through the campus. Such applications can be particularly useful for individuals with visual or hearing impairments, as they can assist them in navigating the campus paths.

Moving forward, continued research and innovation are crucial to maximizing the benefits of NB-IoT technology, while addressing challenges related to its scalability, cost-effectiveness, and integration with emerging technologies.

# FUNDING

The study did not receive funding from any institution.

# DECLARATIONS

## *Conflict of Interest*

The researchers declare no conflict of interest in this study.

## *Informed Consent*

This study does not require informed consent because it focuses on the assessment of a wireless communications technology. There is no interaction with human subjects, nor is any personal data collected. The research solely analyzes the technical characteristics and performance of the technology itself using accessible research publications and papers.

## *Ethics Approval*

The study involves information freely available via open-access journals and websites accessible by the public. It also does not involve the health and well-being of humans and animals; hence an Ethics Approval is not applicable.

*Use of Generative AI And AI-Assisted Technologies in the Writing Process*

During the preparation of this work the authors used QuillBot to improve paraphrasing, Jenni.ai for text generation and consolidated reference citation, elicit.org for simplified related literature review, and gptzero.me for GPT detection and control. After using these tools and services, the authors reviewed and edited the content as needed and took full responsibility for the content of the publication.

Preethy, A., & Parthasarathy, R. (2021). A Quantitative Research Analysis on the Impact of Smart Classroom with IoT Adoption on Students' Academic Performance in Higher Education Programs. *High Technology Letters*, 27, 825–839.

Ramani, P., Palanisamy, A. P., Sekhar, V., & Kumar, N. S. (2023). Smart Attendance Monitoring System Using IoT. In *2023 9th International Conference on Advanced Computing and Communication Systems (ICACCS)*, 1099–1104. https://doi.org/10.1109/icaccs57279.2023.10112850

Rastogi, E., Saxena, N., Roy, A., & Shin, D. R. (2020). Narrowband Internet of Things: A Comprehensive Study. *Computer Networks*, 173, 107209. https://doi.org/10.1016/j.comnet.2020.107209

RF & Wireless Vendors and Resources. (n.d.). *NB-IoT Architecture | LTE-NB IoT Architecture*. RF Wireless World. Retrieved March 30, 2024, from https://www.rfwireless-world.com/Terminology/LTE-NB-IoT-Architecture.html

Sri Madhu, B. M., Kanagotagi, K., & Devansh. (2017). IoT-based Automatic Attendance Management System. *2017 International Conference on Current Trends in Computer, Electrical, Electronics and Communication (CTCEEC)*, 83–86. https://doi.org/10.1109/CTCEEC.2017.8455099

Vos, G. (n.d.). *What is Narrowband IoT (NB-IoT)? (2021 Update) | Sierra Wireless*. Retrieved March 30, 2024, from https://blog.sierrawireless.com/what-is-nb-iot

Yasmin, R., Pouttu, A., Mikhaylov, K., Niemela, V., Arif, M., & Liinamaa, O. (2020). NB-IoT Micro-Operator for Smart Campus: Performance and Lessons Learned in 5GTN. *2020 IEEE Wireless Communications and Networking Conference (WCNC)*, 1–6. https://doi.org/10.1109/WCNC45663.2020.9120621
**Author's Biography**

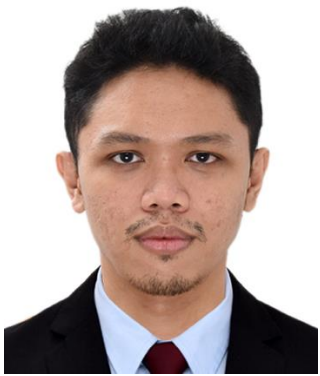

Lyberius Ennio Fenix Taruc, PCpE, is a master's degree student of Computer Engineering with a specialization in Data Science & Engineering at the Polytechnic University of the Philippines – Manila. He finished his bachelor's degree in Computer Engineering (2011) at the Far Eastern University – Institute of Technology, Philippines, where he is also a part-time faculty, in the Computer Engineering Department. A practicing Professional Computer Engineer, PCpE (2023), he also works as a Manager in Globe Telecom, Inc., under the Core Network Planning and Engineering (CNPE) of the Network Technical Group (NTG). His current role in Globe Telecom focuses more on the planning and engineering of cloud infrastructure used by mobile broadband core services. His research interests include Machine Learning, specifically Natural Language Processing, Network Functions Virtualization (NFV), and Cloud Computing.

3056

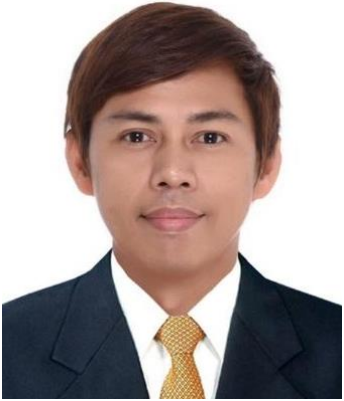

Arvin Roxas De La Cruz, PhD, PCpE, is the Graduate School Research and Extension Chief, one of the Associate Professors of the Graduate School, Open University – Graduate School and College of Engineering, and Program Chair of MS Computer Engineering Program specializing in Data Science & Engineering and Apply Cybersecurity & Digital Forensics at PUP Manila. His completed postgraduate studies include a Doctorate in Management, a Master's in Management Engineering, and a Master's in Information Technology. He serves as an Academic Head across several educational institutions and has served as a consultant for schools in ICT, Business, Engineering, and TESDA programs across the country. His research interests include Management Engineering, Machine Learning and Data Science & Engineering, Cybersecurity & Digital Forensics, Embedded Systems, and Modeling with Simulation Systems.